\newcommand{\be}{\begin{equation}}
\newcommand{\ee}{\end{equation}}
\newcommand{\bea}{\begin{eqnarray}}
\newcommand{\eea}{\end{eqnarray}}

\documentclass[prb,amssymb,showpacs]{revtex4}
\usepackage{graphicx}
\usepackage{dcolumn}
\usepackage{bm}
\usepackage{amsbsy}
\usepackage{amsmath}
\usepackage{amsfonts}

\begin{document}
\title{Effect of inter-wall surface roughness correlations on optical
spectra of quantum well excitons}
\author{I. V. Ponomarev}
\email{ilya@physics.qc.edu}
\author{L. I. Deych}
\author{A. A. Lisyansky}
 \affiliation{Department of Physics, Queens College of the City University
of New
 York,\\
Flushing, NY 11367}
\date{\today }

\begin{abstract}
We show that the correlation between morphological fluctuations of
two interfaces confining a quantum well  strongly suppresses a
contribution of interface disorder to inhomogeneous line width of
excitons. We also demonstrate that only taking into account these
correlations one can explain all the variety of experimental data
on the dependence of the line width upon thickness of the quantum
well.
\end{abstract}
 \maketitle



\section{\label{sec:sec1}Introduction}

Absorption and luminescence exciton spectroscopy are one of the
most important tools for studying quantum wells (QW) as well as
other semiconductor heterostructures. Therefore, one of the most
fundamental problems in physics of these systems is establishing
connections between spectral line shapes and microscopic
properties of the respective structures. A great deal of efforts
was devoted to this problem over the last half of century, and it
has been established, among other things, that at low temperatures
the exciton line width in absorption and photoluminescence spectra
in quantum wells is predominantly
inhomogeneous\cite{EfrosRaikh88}. The shape of the spectra in this
case is determined by various types of disorder present in a
structure, and it is currently generally accepted that the
spectral width in QW is directly related to the quality of
interfaces, so that the luminescence spectra provide a quick and
simple quality assurance tool for QW growth\cite{Weisbuch81a}.

However, as it will be seen from the subsequent analysis, in spite
of all the efforts, the existing theories of inhomogeneous
broadening of excitons in QW still cannot satisfactory explain all
the diverse experimental data collected in this area. While it is
quite plausible that the present theories suffer from several
drawbacks, it is the goal of this paper to point out at one of the
important circumstances completely overlooked in the previous
studies. We will show here that it is not possible to explain
experimental results collected by different groups for different
material systems without taking into account so called
\textit{inter-wall correlations}, which describe the fact that
properties of two interfaces forming a quantum well are
correlated, especially for narrower wells. While direct
morphological analysis done with cross-sectional scanning tunnel
microscopy\cite{Yayon02a}, scattering ellipsometry\cite{Germer00a}
and X-ray reflection measurements,\cite{Holy94a,Kondrashkina97}
give convincing experimental evidences of such correlations, their
role in optical spectra of QW was never studied before. Another
relevant example of such correlations is vertical stacking of
quantum dots\cite{Bimberg99}, where vertical correlation length is
observed up to the $80\,$ monolayer thickness.

One of the main objectives of this paper is to show that the
inter-wall correlations can significantly modify optical spectra,
and that taking these correlations into account is necessary in
order to achieve even a qualitative agreement between theory and
experimental results. One of the important practical conclusions
of this work is that narrow lines not always mean a good quality
interfaces contrary to the popular belief, but can be the result
of a line narrowing effect of the inter-wall correlations.

The paper is organized in the following way. In the next section
we shall provide a brief review and critical analysis of existing
experimental results and relevant theories. In subsequent section
we will generalize the earlier theories of the interface
disorder\cite{Weisbuch81a,Zimmermann97a,Zimmer92a,Srinivas92a,Castella98}
to include effects of the vertical correlations. In the last
section we will compare the results of our analysis with
experiments. The paper is concluded by an Appendix, in which we
comment on the role of the lateral (in-plane) correlation length
in the optical spectra of QW.

\section{Comparison between current theories and experiment:
critical review}

Since a QW is generally a heterostructure formed by a binary
semiconductor ($AB$) and a ternary disordered alloy
($AB_{1-x}C_x$), there are two types of disorder responsible for
the inhomogeneous broadening. One is compositional disorder caused
by concentration fluctuations in a ternary component of the QW as
well as random diffusion across the
interface\cite{EfrosBaranov78,EfrWetWor95}. The other source of
inhomogeneous broadening in QWs is associated with the roughness
of the interface caused by the formation of monolayer islands at
the interfaces resulting in local changes in the well
thickness.\cite{Weisbuch81a,Zimmermann97a,Zimmer92a,Srinivas92a,Castella98}
The quality of the interfaces is very sensitive to the ambient
parameters of the growth process. Depending on growth conditions,
the atoms deposited on the surface can form ``islands" of various
lateral sizes with different correlation scales. These
morphological changes manifest themselves in the shape and width
of optical spectra of QW excitons. Since both types of disorder
can be ultimately traced to local changes in concentration, an
accurate distinction between them is not a trivial task, and was
first elucidated in Ref.~\onlinecite{Zimmermann97a}.

Theoretical studies of effects due to  compositional and interface
disorder on  absorption and photoluminescence spectra have a long
history (for review articles see, for example,
Refs.~\onlinecite{EfrosRaikh88,Herman91a,Runge02a} and references
therein.). This problem can be divided into two fairly independent
parts. The first  deals with the derivation of the random
potentials acting on excitons in QW's from the properties of
microscopic fluctuating parameters (concentrations, well
thickness, etc.). Its main objective is to calculate correlation
functions of these potentials. The second part of the problem
consists of calculations of characteristics of excitons subjected
to these potentials and in establishing relations between
characteristics of optical spectra (line width, shape) and
properties of the potentials (r.m.s. fluctuation and correlation
length). Both these problems were carefully studied in the past,
but since the focus of the present work is on the former we shall
discuss it in more detail.

The main object of our discussion is the r.m.s value of the random
potentials, $W$, defined as $W=\sqrt{\langle V_{eff}({\bf
R})^2\rangle}$, and its dependence on the microscopic parameters
of the QW. Here $V_{eff}$ is the effective potential acting on
excitons due to both compositional and interface disorder. Current
theories predict distinctly different properties for contributions
to the potential from these two types of disorder. In particular,
the dependence of $W$ upon the width of the well, $L$, for a
narrow well is predicted as being $\propto L$ for the interface
disorder. For the compositional disorder, it depends on whether
the QW is formed by a ternary alloy or a binary
system\cite{EfrWetWor95}. In the former case (as in
In$_x$Ga$_{1-x}$As/GaAs)  $W\propto L^{3/2}$, while in the case
when the QW is formed by a binary material, the dependence is
$W\propto \sqrt{L}$. In the three-dimensional limit
$L\rightarrow\infty$ the contributions of the two types of
disorder to $W(L)$ are also different. The contribution of the
interface disorder for large $L$ decreases as $1/L^3$, while the
role of the alloy disorder again depends upon the type of the
structure. If QW is formed by a ternary alloy $W$ only weakly, as
$1/\sqrt{L}$, decreases with the width before reaching a constant
three dimensional limit. If, however, the alloy forms barriers,
$W$ decreases with $L$ much faster as $\exp(-\kappa_0L)/L^3$,
where $\kappa_0$ is an inverse penetration length of ground state
wave function in the barrier region. Although in both cases the
function $W(L)$ has a maximum at some intermediate values of $L$,
the position of the maximum, and the shape of the function $W(L)$
differ for the two types of disorder.

\begin{figure}[tbp]
\includegraphics{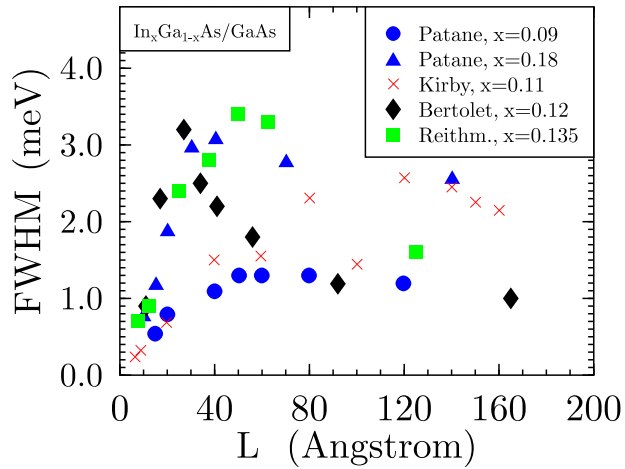}
 \caption{Experimental dependences of the low temperature exciton full width on half maximum (FWHM) on the QW average size, $L$.
 All the data are presented for In$_x$Ga$_{1-x}$As/GaAs QW's. The results were taken from the following
 references: Patan\`{e}, et all.\cite{patane95a}, Kirby, et all.\cite{Kirby89}, Bertolet, et all.\cite{Bertolet88}, Reithmaier, et all.
\cite{Reithmaier91}.} \label{fig_exp}
\end{figure}

While the qualitative picture of disorder induced broadening is
understood rather well, attempts at a quantitative comparison of
theoretical predictions with experiments face significant
difficulties. In Fig.~\ref{fig_exp} we have collected experimental
data for dependencies of low temperature photoluminescence exciton
line widths on QW average thickness, $L$. All the data are for
In$_x$Ga$_{1-x}$As/GaAs heterostructures, and represent
experimental results from  several research
groups\cite{patane95a,Kirby89,Bertolet88,Reithmaier91}. While all
the results show a non-monotonous dependence in accord with
theoretical expectations, the maxima have different positions and,
do not seem to have a regular dependence on concentration: the
peak for $x=0.18$ lies between peaks for $x=0.12$ and $x=0.135$.
The maxima also have different heights and sharpness. For example,
the full-width at half maximum (FWHM) for $x=0.09$ and $x=0.11$
have very smooth behaviors more characteristic for compositional
disorder, while other data have rather sharp features more typical
for interface disorder. Finally, the values of FWHM at large
$L$, which are determined mainly by compositional disorder, 
are scattered over quite a broad range.

Let us analyze Fig.~\ref{fig_exp} in light of present theories.
Regarding the alloy contribution to the line width, we note that
in quantum mechanical approaches to this
problem\cite{EfrosBaranov78,Lyo93a,EfrWetWor95,Zimmermann97a} this
contribution is completely determined by alloy concentration and
QW width; the theories do not contain any unknown parameters that
could be used to fit theoretical predictions to the experiments.
Therefore, one can  directly compare theoretical predictions with
experimental results using the large $L$ asymptote of the line
width. Initial calculations for alloy induced disorder in the
bulk\cite{EfrosBaranov78,Efros83a} and in the quantum
wells\cite{EfrWetWor95} were done in the adiabatic approximation,
where Bohr's radius of an exciton, $a_B$, was assumed to be much
larger than a correlation length of the effective potential,
$\ell_c$. If one applies that theory\cite{EfrWetWor95} to the case
of an In$_{0.12}$Ga$_{0.88}$As/GaAs QW, the FWHM comes out to be
$=0.36$ meV for $L=150$ $\AA$. This value is much smaller than the
observed bulk value\footnote{In fact, the value calculated with
the help of Eq. (A4) in Ref.~\onlinecite{EfrWetWor95} would be
even smaller, since the authors omitted without any reasons the
cross-term $2\alpha_e\alpha_h\int\chi_e^2\chi_h^2 dz$ which is of
the same order of value as other two terms.}. The same order of
magnitude results are obtained in the semi-classical limit of the
theory\cite{Singh85a,Zimmermann90a} with the Gaussian shape of the
exciton line width. One can reasonably argue\cite{EfrosRaikh88}
that the adiabatic approximation fails for a heavy hole and a
light electron ($m_h\gg m_e$) both being subject to short-range
energy fluctuations. Even for underlying ``white-noise" disorder,
the effective potential felt by an exciton has two different
correlation lengths. A rather massive hole will average only the
small volume around center-of-mass (COM) of the exciton, whereas a
light electron is spread out over a much larger area of the order
of $a_B^2$. As a result, the hole will be much more sensitive to
the compositional fluctuations, and its contribution to the
effective disorder potential will be enhanced by the factor
$(M/m_e)^2$, where $M=m_e+m_h$. However, the line widths found
using improved theories\cite{Lyo93a,Zimmermann97a} turn out to be
much larger than the experimental results (see also
Fig.~\ref{fig_fit} below). Thus, existing theories  cannot produce
an accurate result for the alloy disorder contribution to the
exciton line width.


It is less straightforward to compare the theory with experiments
for interface roughness induced broadening  because it is
difficult to separate contributions from two types of disorder in
the regime of small and intermediate values of $L$. One could hope
to identify the most important contribution by the slope of the
$W(L)$ dependence at small $L$, but, unfortunately, the accuracy
of the existing data does not allow one to distinguish between $L$
or $L^{3/2}$ dependencies. The manifestation of the interface
disorder in optical spectra is clearer in the case of systems
where growth interruption resulted in formation of sufficiently
large monolayer islands. It was shown in
Ref.~\onlinecite{Castella98} that these islands could be
responsible for the observed splitting of the exciton spectra.
Also, interface roughness has been studied directly by such
experimental techniques as microphotoluminescence,
cathodoluminescence, transition electron microscopy, or scanning
tunnelling microscopy.
\cite{Weisbuch81a,patane95a,Kirby89,Bertolet88,Reithmaier91,Srinivas92a,
Zhao02a,Bimberg86a,Ourmazd89a,Christen89a,Herman91a}

The diversity of experimental behavior for FWHM shown in
Fig.~\ref{fig_exp} presents a significant difficulty for existing
theories of interface contribution to the line
width\cite{Singh84a,Zimmermann97a} even from the point of view of
qualitative interpretation of the results. Indeed, the statistical
properties of an interface are usually characterized by two length
parameters: the thickness fluctuation, $h$, and the lateral
(in-plane) correlation length, $\sigma_{\perp}$. With some
reservations they are often taken to be equivalent to the average
height and the lateral size of the islands at the interface. The
size of height fluctuations, $h$, usually has  a very restricted
range of variations of one or two monolayers. In the
semi-classical limit of the Gaussian shape line width, the
broadening is proportional to the product $h\sigma_{\perp}$, which
is really only one flexible parameter of the theory. With the help
of this parameter it is possible to adjust the relative heights of
the maxima but not their positions and particularly the sharpness
of the peak. Even if one takes into account contributions from
both types of disorder, it is still not possible to explain the
variations between the optical spectra of different samples, a
problem addressed in the last section of the paper.

It is clear from the provided analysis that existing theories of
inhomogeneous broadening of excitons are unable to quantitatively
explain experimental data. We suggest in this paper that one of
the reason for this failure is the neglect of inter-wall
correlations mentioned in the Introduction. While this idea does
not fix the problems caused by wrong estimates of alloy disorder
contribution, we will show in the subsequent sections of the paper
that it does allow to explain all variety of experimental results
related to the properties of the curves in Fig.~\ref{fig_exp} in
the vicinity of their maxima.

\section{Statistical properties of the interfaces and exciton effective potential\label{sec:2}}
\subsection{A model of the interface disorder }
\begin{figure}[tbp]
\includegraphics{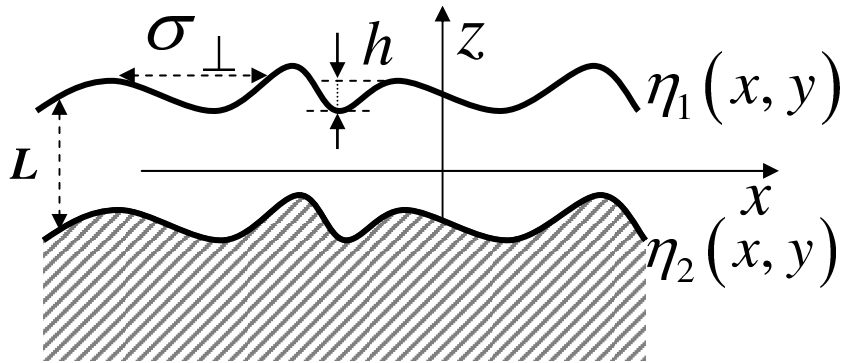}
 \caption{Characteristic length scales describing interface roughness of a QW.} \label{fig2}
\end{figure}

 In order to make the problem tractable, we introduce standard simplifications
assuming that both  conduction and valence bands are
non-degenerate, and that they both have an isotropic parabolic
dispersion characterized by the masses $m_e$ and $m_h$,
respectively.  Throughout the paper we use effective atomic units
(a.u.), which means that all distances are measured in units of
the effective Bohr radius $a_B=\hbar^2\epsilon/\mu^*e^2$, energies
in units of $E_{a.u.}=\mu^*e^4/\hbar^2\epsilon^2\equiv 2\textrm{
Ry}$, and masses in units of reduced electron-hole mass $\mu^*$,
where $1/\mu^*=1/m_e^*+1/m_h^*$. In this notation
$m_{e,h}=m_{e,h}^*/\mu^*$, where $m_{e,h}^*$ are effective masses
of an electron and a heavy hole. We will choose $z$-axes in the
direction of growth of the structure (vertical direction). The
plane perpendicular to this direction is the lateral plane (see
Fig.~\ref{fig2}).
We measure electron and hole energies in the QW from the
conduction and valence band edges of the barrier respectively.
Then the potential of a QW with interface roughness is given by
\begin{equation}\label{QWpot_tot}
  U_{e,h}(\textbf{r})=-V_{e,h}\left[\theta(z+L/2-\eta_1(x,y))-\theta(z-L/2-\eta_2(x,y))\right]\approx
U_{e,h}^{(0)}(z)+\delta U^{intf}_{e,h}(\textbf{r}),
\end{equation}
where $\theta(z)$ is a step-function, $V_{e,h}$ are differences in
off-set band energies, and
\begin{eqnarray}
U_{e,h}^{(0)}(z)&=&-V_{e,h}\left[\theta(z+L/2)-\theta(z-L/2)\right],\label{UQW0}\\
\delta U^{intf}_{e,h}(\textbf{r})& =&
V_{e,h}\left[\eta_1(x,y)\delta(z+L/2)-\eta_2(x,y)\delta(z-L/2)\right].\label{Uinterf}
\end{eqnarray}
Random functions $\eta_{1,2}(x,y)$, with zero mean, characterize
the deviation of the $i$th interface from its average position.
The perturbation expansion of the $\theta$-function is justified
because an interface roughness is almost always small for typical
parameters in semiconductor heterostructures. The presence of two
functions, $\eta_{1,2}(x,y)$, distinguishes Eq.(\ref{UQW0}) and
Eq.(\ref{Uinterf}) from the respective equations of
Ref.\onlinecite{Zimmermann97a}, where the roughness of only one
interface was taken into account.

The statistical properties of the interfacial roughness in
multilayered systems can be characterized by the height-height
correlation functions
\begin{equation}
\langle \eta_i(({\bm{\rho}}_1) \eta_j(({\bm{\rho}}_2)\rangle =
h^2f_{ij}\zeta(|{\bm{\rho}}_1-{\bm{\rho}}_2|),
\end{equation}
where $h$ is an average height of interface inhomogeneity, and
$\langle \ldots \rangle$ denotes an ensemble average. We assume
here that the dependence of both diagonal and non-diagonal
correlations on the lateral coordinates ${\bm \rho}$ is described
by the same function $\zeta({\bm \rho})$.   For diagonal elements
$f_{ii}\equiv 1$, and the respective  functions  describe lateral
correlation properties of a given interface
(\emph{self-correlation functions}). Non-diagonal elements with
$i\neq j$ introduce correlations between different interfaces; the
respective quantity $f_{12}(L/\sigma_{\|})$, which can be called a
\emph{cross-} or \emph{vertical-correlation function}, is a
function of the average width of the well and is characterized by
the vertical correlation length $\sigma_{\|}$ (a subscript ``$\|$"
denotes that the direction of the vertical correlation is parallel
to the direction of growth).

The effect of the inter-wall vertical correlations has been
previously considered in studies of conductivity of thin metallic
films\cite{Meyerovich99,pon02a}. To the best of our knowledge, in
all previous microscopic studies of the exciton line shape in
optical spectra of QWs, these correlations were omitted. Such an
approximation is valid for wide QWs, but in the case of narrow QWs
the vertical correlations are experimentally
confirmed\cite{Yayon02a,Germer00a,Holy94a,Kondrashkina97,Bimberg99}
and should be taken into account. In the limit
$L/\sigma_{\parallel}\ll 1$ it is reasonable to assume that the
inter-wall correlation function $f_{12}$ tends to unity, which
means that for the very small separation between the interfaces
one random surface spatially repeats the pattern of the other.
(See Fig.~\ref{fig2}). As we shall see below, the effect of
interface disorder in this case tends to cancel out at least in
the first order of the perturbation theory. For the sake of
concreteness we will assume below that
\begin{equation}\label{figform}
  f_{12}=\exp(-L^2/\sigma_{\parallel}^2).
\end{equation}
The value of the vertical correlation length  $\sigma_{\bot}$
depends on the growth process. In the following analysis we will
also assume the Gaussian form for the lateral correlation
function:
\begin{equation}\label{gausscor}
  \zeta({\bf R})=\exp\left(-R^2/2\sigma_{\perp}^2\right).
\end{equation}
The limit $\sigma_{\perp}\rightarrow 0$ corresponds to the
``white-noise" correlator
\begin{equation}\label{whitenoise}
  \zeta({\bf R})=2\pi\sigma_{\perp}^2\delta({\bf R}).
\end{equation}

Following the standard procedure\cite{EfrosBaranov78} described in
numerous papers we derive the Schr\"{o}dinger equation for the
center of mass (COM) exciton motion subjected to an effective
random potential,
\begin{equation}\label{SEforCOM}
\left[-\frac{\triangle_{\textbf{R}}}{2M}
+U_{eff}(\textbf{R})\right]\psi_i(\textbf{R})=\varepsilon_i\psi_i(\textbf{R}),
\end{equation}
with $U_{eff}(\textbf{R})$ given by
\begin{equation}\label{Ueff01}
  U_{eff}(\textbf{R})=\int\left( \delta U_e+\delta
  U_h\right)\phi^2({\bm{\rho}})\chi^2_e(z_e)\chi^2_h(z_h)\;
  d^2\rho dz_e dz_h\equiv U_e(\textbf{R})+U_h(\textbf{R}).
\end{equation}
where $\delta U_{e,h}$ are defined in Eq.~(\ref{Uinterf}),
${\bm{\rho}}={\bm\rho}_e-{\bm\rho}_{h}$, and
$\textbf{R}=(m_e{\bm\rho}_{e}+m_h{\bm\rho}_{h})/M.$
 For symmetric QW, $\chi(-L/2)=\chi(L/2)$, and from Eq.~%
(\ref{Uinterf}) we obtain
\begin{equation}\label{Ueff04}
U_{e,h}(\textbf{R})=V_{e,h}\chi_{e,h}^2(L/2)\int\left[
\eta_1(\textbf{R}\pm\beta_{h,e}\bm{\rho})-\eta_2(\textbf{R}\pm\beta_{h,e}{\bm\rho})\right]
\phi^2({\bm\rho})\;d^2\rho.
\end{equation}
Here $\beta_{h,e}=m_{h,e}/M$. Hereafter we will omit an explicit
dependence on $L$ for the electron and hole wave functions
$\chi_{e,h}^2(L/2)$ always assuming that their values are taken at
the position of the interface. The correlation function for the
effective potential $U_{eff}({\bf R})$ can then be expressed as
\begin{equation}
\langle U(\mathbf{R}_1)U(\mathbf{R}_2)\rangle\equiv
T_{ee}+T_{hh}+2T_{eh},
\end{equation}
where
\begin{eqnarray}
T_{ii}&=&2h^2V_{i}^2\chi_{i}^4\left(1-f_{12}(L)\right)\int\;d^2\rho
d^2\rho'\phi^2(\rho)\phi^2(\rho')
\zeta\left(|\textbf{R}-\beta_j({\bm\rho}-{\bm\rho'})|\right),
\label{T1Eq}\\
T_{eh}&=&2h^2V_{e}V_{h}\chi_{e}^2\chi_{h}^2\left(1-f_{12}(L)\right)\int\;d^2\rho
d^2\rho'\phi^2(\rho)\phi^2(\rho')
\zeta\left(|\textbf{R}-\beta_h{\bm\rho}+\beta_e{\bm\rho'}|\right),
\end{eqnarray}
where  $i,j=e$ or $h$ and $\textbf{R}=\textbf{R}_1-\textbf{R}_2$.

 In all these expressions, the terms in front of the integrals
determine the dependence of the correlation function on QW width
$L$, while the integrals themselves determine the spacial
correlations in lateral dimensions. Each of these terms can be
presented in a form
\begin{eqnarray}
  T_{ij}(L,R) & = & 2h^2V_iV_jF_{ij}(L)G_{ij}(R),\label{T1form}\\
  F_{ij}(L)   & = &\chi_i^2\chi_j^2\left(1-f_{12}(L)\right),\label{T2form}\\
  G_{ij}(R)   & = &
\int\;d^2\rho d^2\rho'\phi^2(\rho)\phi^2(\rho')
\zeta\left(|\textbf{R}-\beta_j{\bm\rho'}+\beta_i{\bm\rho}|\right),\label{T3form}
\end{eqnarray}
where indexes $i,j$ take values $e$ and $h$ (note that the order
of these indices in the integrand is important).
\subsection{Dependence on well thickness}
\begin{figure}[tbp]
\includegraphics{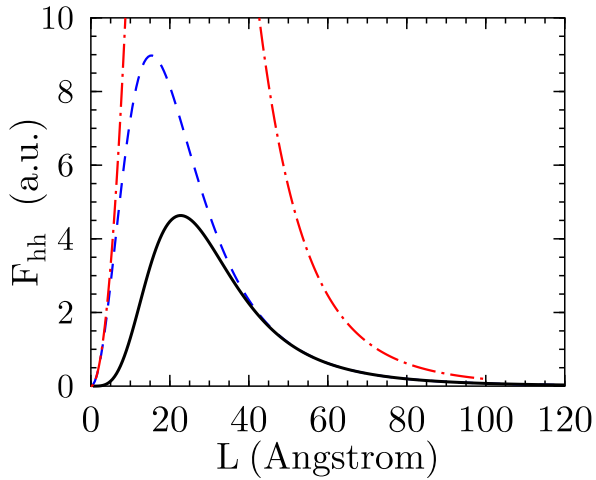}
 \caption{Dependence of the correlation function of the hole-hole interface roughness potential
 on the well width in an In$_{0.12}$Ga$_{0.88}$As/GaAs QW. The solid line is the function
 $F_{hh}(L)$. The dashed line is the fourth degree of the hole
 wave function at the interface. The dotted-dashed line shows approximations of this function for small and large
 $L$, given by Eqs.~(\ref{chi_largeL}) and (\ref{chi_smallL}).
 The maximum of $\chi_h(L)^4$
is approximately located at $1/\kappa_{0h}=1/\sqrt{2m_hV_h}$.}
\label{fig2b}
\end{figure}

 Let us first focus on
the functions $F_{ij}(L)$ which determine the dependence of the
total correlation function on QW width $L$. For the sake of
concreteness we consider in detail  function $F_{hh}$. The
analysis of other functions is similar. The dependence of $F_{hh}$
on $L$ comes from two factors. The first factor is the fourth
degree of the electron QW wave function, $\chi_{h}^4$, calculated
at the interfaces, $z=\pm L/2$.
 This dependence in In$_{0.12}$Ga$_{0.88}$As/GaAs QW is shown in Fig.~\ref{fig2b} by a dashed line.
It is easy to understand the behavior of $\chi_h^4$ for the cases
of large and small width. The characteristic scale here is given
by $1/\kappa_{0h}=1/\sqrt{2m_hV_h}$, since this scale determines
the number of energy levels in a finite QW ($N=1+\left[\kappa_h
L/\pi\right]$).

For finite QW the ground state wave function has a piece-wise form
\begin{equation}
\chi\left(  z\right) =\left\{
\begin{array}
[c]{cc}%
A\cos(kz), & z\leq |L/2|\\
B\exp(-\kappa z), & z\geq |L/2|,
\end{array}
\right. \label{chiQW}
\end{equation}
where $\kappa=\sqrt{\kappa_0^2-k^2}$ and $k$ is the ground state
wave vector. From the normalization and matching conditions, one
can readily obtain the value of  square of the wave function at
the interface
\begin{equation}\label{ae2SE}
\chi^2=B^2\exp(-L)=\left[\frac{\kappa}{1+\kappa^2/k^2+\kappa
L\kappa_0^2/(2k^2)}\right].
\end{equation}
 In the case of a wide QW, there are many discrete levels in
the well, and the well is ``almost infinite": $k \approx
\pi/L-2\pi/L^2\kappa_0$, $ \kappa \approx  \kappa_0,$ and
\begin{equation}
\chi^2 \approx \frac{2\pi^2\kappa_0}{\kappa_0^3 L^3
 +2\kappa_0^2 L^2+2\pi^2}\sim\frac{1}{L^3}. \label{chi_largeL}
\end{equation}
Since for large well widths the vertical inter-wall correlations
are negligible, $f_{12}\rightarrow 0$, Eq.~(\ref{chi_largeL})
determines the total decrease of the interface correlator with
increasing width.

 In the opposite case of very narrow QWs, there is only one shallow
level, which can be determined from the $\delta$-functional
potential approximation: $k \approx  \kappa_0-L^2\kappa_0^3/8$,
$\kappa \approx  \kappa_0^2 L/2$, and
\begin{equation}\label{chi_smallL}
\chi^2 \approx \frac{\kappa_0^2 L}{2}.
\end{equation}
If one neglects vertical correlations, this expression describes
the suppression of the interface disorder for narrow wells because
of a decreased portion of the hole (electron) wave function inside
the well. This result was obtained earlier in, for instance,
Ref.\onlinecite{Singh85a}. Inter-wall correlations, however,
significantly modify this dependence. For lengths smaller than the
vertical correlation length $\sigma_{\|}$ we have
\begin{equation}\label{intw_sm}
  1-f_{12}\sim\left(\frac{L}{\sigma_{\parallel}}\right)^{\gamma},
\end{equation}
where $\gamma$ is determined by the form of the inter-wall
correlation factor $f_{12}$. For example, for Gaussian or
Lorentzian dependences of $f_{12}(L)$, the parameter  $\gamma=2$,
while for the exponential form of this function, $\gamma=1$. Thus
we obtain that, in narrow QWs, interface correlations are strongly
suppressed by the factor
\begin{equation}\label{Fmain}
  F_{ij}(L)\sim L^{2+\gamma},\qquad L<\sigma_{\parallel},1/\kappa_0.
\end{equation}
While the transition between two asymptotic behaviors of $\chi$ is
determined by the parameter $\kappa_0$, the behavior of $f_{12}$
depends on the correlation length $\sigma_{\|}$, which is a
completely independent parameter. Experimental data suggest that
it is quite possible for $\sigma_{\|}$ to be much larger then
$\kappa_0$. In this case, inter-wall correlations can affect not
only the $L\rightarrow 0$ asymptotic of $W$, but also its behavior
at $L\gg\kappa_0$. Instead of $1/L^3$ behavior one would have a
much slower decrease of $W$ with $L$: $W\propto L^{\gamma-3}$.

From the behavior of the wave function (\ref{ae2SE}) we can see
that the maximum of $F_{hh}(L)$ is reached in the vicinity of
$L\sim 1/\kappa_{0h}$. However, as the previous analysis
demonstrates, the vertical correlation function $f_{12}$ can
significantly shift this position; it can also change the height
and shape of the peak. Thus, the presence of the inter-well
correlation term $1-f_{12}$ in function $F_{hh}$ can naturally
explain all variety of experimental results related to the
properties of the curves in Fig.~\ref{fig_exp} in the vicinity of
their maxima. The graph of function $F_{hh}$ with inter-wall
correlations taken into account is shown in Fig.~\ref{fig2b}. One
can see that these correlations indeed significantly affect the
shape of this function. While we only discussed properties of
hole-hole correlator, it is clear that behavior of the
electron-electron and hole-electron terms is similar.
\subsection{Dependence on lateral correlation length}
In order to investigate the results of the interplay between
lateral and inter-wall correlations, let us now analyze the
lateral correlation functions $G_{ij}(R)$. Their behavior is
determined by the ratio of the potential correlation length,
$\sigma_{\perp}$ to the average size of the exciton in a plane, as
well as by dimensionless parameters $\beta_e$ and $\beta_h$. In
order to evaluate the respective integrals, we chose the
normalized ground state function of exciton relative motion in a
quasi-two-dimensional form\cite{Bastard82a}
\begin{equation}\label{excwf}
\phi(\rho)=\sqrt{\frac{2}{\pi\lambda^2}}\exp(-\rho/\lambda),
\end{equation}
where $\lambda$ is a variational parameter that indicates the
average exciton size.

For the ground state function, Eq.~(\ref{excwf}), and the
height-height correlation function, Eq.~(\ref{gausscor}), the
lateral dependence of the correlator $G_{ij}(R)$ can be presented
as a function of two parameters $G_{ij}(R)\equiv
G_{ij}(R;y_j,\alpha)$, where
\begin{equation}\label{COMLcor01}
y_i=\frac{\sqrt{2}\sigma_{\perp}}{\beta_i\lambda},
\end{equation}
and $\alpha=\min{(\beta_i,\beta_j)}/\max{(\beta_i,\beta_j)}$.
Parameter $\alpha$ is equal to unity for the electron-electron and
the hole-hole correlator. For the cross term $G_{eh}$ this
parameter is equal to $m_e/m_h$, which is much less than unity for
the majority of semiconductor materials. This fact allows for
additional simplifications when evaluating the integrals. The
parameter $y_{e(h)}$ defines the ratio of the renormalized lateral
correlation length $\beta_{e(h)}\lambda$ of the effective hole
(electron) potential to the original correlation length of the
interface fluctuations. For holes with $m_h\gg m_e$ this
renormalized correlation length is much smaller than the
corresponding length for the electrons. The latter implies that
the interface disorder has a bigger impact on the holes compared
to the electrons. This is also true for the contribution of alloy
disorder to the broadening\cite{EfrosBaranov78}. Thus, the lateral
correlation function can be rewritten in the following form:
\begin{equation}\label{COMLcor02}
G_{ij}(R;y_j,\alpha)=\frac{4y_j^4}{\alpha^2\pi^2} \int\;d^2\rho
d^2\rho'\exp\left[-2y_j(\rho'+\alpha^{-1}\rho)\right]
\exp\left(-|\textbf{R}-\bm{\rho'}+\bm{\rho}|^2\right).
\end{equation}
We shall focus on its value at the origin, $G_{ij}(0;y_j,\alpha)$,
which determines the variance of the potential, $W$. The
expression for $W$ in this case can be presented in the following
form:
\begin{equation}\label{U02_gauss}
W^2=2h^2\left[1-f_{12}(L/\sigma_{\bot})\right] \left[V_h^2\chi_h^4
G_{hh}(0;y_e,1) + 2V_eV_h\chi_e^2\chi_h^2G_{eh}(0;y_h,\alpha)
 + V_e^2\chi_e^4G_{ee}(0;y_h,1)\right].
\end{equation}

\begin{figure}[tbp]
\includegraphics{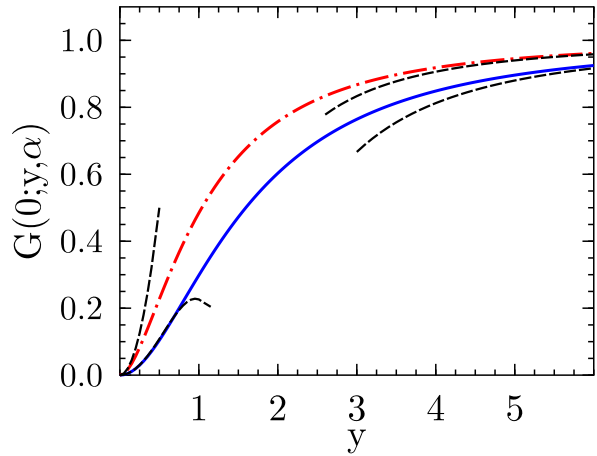}
 \caption{Thick solid line is the lateral correlation function $G(0;y,1)$. The dashed lines are its
 asymptotes given by Eq.~(\ref{COMLcor04c}). The thick dotted-dashed line is a
 cross-term correlator $G(0;y,0)$. Its asymptotes (dashed lines) are given by Eq.~(\ref{COMLcor03b})}\label{figCor}
\end{figure}

The calculation are  easier for the cross-term $G_{eh}$, because
we can take advantage of smallness of $\alpha\ll 1$:
\begin{equation}\label{COMLcor03}
G_{eh}(0;y_h,\alpha)= 4y_h^2 \int dt\, t\exp(-t^2-2y_h
t)+O(\alpha^2)\\
=2y_h^2/(1+\alpha)^2-2y_h^3\exp(y_h^2)\sqrt{\pi}(1-{\rm
erf}(y_h))+O(\alpha^2),
\end{equation}
where ${\rm erf}(y)$ is the error function. The function
(\ref{COMLcor03}) is shown in Fig.~\ref{figCor}. It has the
following behavior for small and large $y$:
\begin{equation}\label{COMLcor03b}
G_{eh}(0;y_h,0)\approx \left\{
\begin{array}
[c]{cc}%
\displaystyle{\frac{4\sigma_{\perp}^2}{\lambda^2}}  -2\sqrt{\pi}y_h^3+\cdots, & y_h\ll 1,\\
1-\displaystyle{\frac{3}{2y_h^2}}, & y_h\gg 1.
\end{array}
\right.
\end{equation}
The calculations are more cumbersome for the electron-electron and
the hole-hole contributions $G_{ii}(0,y_j,1)$. It is convenient
perform a transformation to the new set of coordinates which
reflect the symmetry of the integral\cite{Bethe}:
\begin{equation}\label{COML04a}
  s=\rho_1+\rho_2,\quad t=\rho_1-\rho_2, \quad
  u=\rho_{12}=\sqrt{\rho_1^2+\rho_2^2-2\rho_1\rho_2\cos\theta}.
\end{equation}
After this transformation\footnote{For detailed derivation of the
volume element in two-dimensional case see Appendix B in
Ref.~\onlinecite{pon99a}} the calculations for $G_{ii}$ are
reduced to the one-dimensional integral
\begin{equation}\label{COML04b}
  G_{ii}(0,y_j,1)=y_j^4\int_0^{\infty} ds\; e^{-2y_js}
  \left[s + \left(2\,s^2-1 \right)e^{-s^2}
  \frac{\sqrt{\pi }\,\mathrm{erf}(is)}{2i}\right].
\end{equation}
 The term in square brackets in the integrand of Eq.~%
(\ref{COML04b}) is a smooth function that behaves as $\sim
(8/3)s^3$ for small $s$, and changes its behavior to $\sim 2s$ for
$s>1$. These dependencies allow one to estimate the asymptotic
behavior for the correlator $G_{ii}(0;y_j,1)$:
\begin{equation}\label{COMLcor04c}
G_{ii}(0;y_j,1)\approx \left\{
\begin{array}
[c]{cc}%
\displaystyle{\frac{y_j^2}{2}}-\frac{23y_j^4}{84}, & y_j\ll 1\\
1-\displaystyle{3/y_j^2}, & y_j\gg 1.
\end{array}
\right.
\end{equation}
The first terms in the series expansions of Eqs.~%
(\ref{COMLcor03b}) and (\ref{COMLcor04c}) for small $y$ correspond
to the white-noise limit of the height-height correlation
function, Eq.~(\ref{whitenoise}), and were obtained previously in
Ref.~\onlinecite{Zimmermann94a}. They can be readily derived by
substitution of a $\delta$-function instead of the last exponent
in the integrand of Eq. (\ref{COMLcor02}). The consecutive terms
in these formulas describe deviations from the white-noise model
in the case of the short-range correlations. The dependence of
correlators $G_{ij}(0;y,\alpha)$ on the respective parameters
$y_i$  is shown in Fig.~\ref{figCor}, from which one can see that
corrections to the white-noise approximation become significant
even at relatively small values of the correlation length,
$\sigma_{\perp}$.

For the ``white-noise" interface roughness, Eq.~%
(\ref{whitenoise}), the analytical results can be obtained for the
more elaborate exciton ground state trial function\cite{pon03a}
\begin{equation}\label{excwf02}
\phi(\rho)=\frac{2\exp(\gamma)}{\sqrt{2\pi\lambda^2(1+\gamma)}}\exp\left(-\sqrt{\rho^2/\lambda^2+\gamma^2/4}\right),
\end{equation}
which more accurately takes into consideration the
three-dimensional character of the exciton. The parameter
$\gamma=2d/\lambda$ determines the ratio of the finite average
distance $d$ between the electron and the hole in the QW to the
two-dimensional Bohr radius $\lambda$. In this case, one can
obtain for the most important hole-hole correlator ($\alpha=1$)
the following expression
\begin{equation}\label{Gij02b}
 G(0,\sigma_{\|},\{\lambda,\gamma\},\beta,1)=
\frac{\sigma_{\|}^2}{\beta^2\lambda^2}\frac{1+2\gamma}{(1+\gamma)^2}.
\end{equation}
This result formally coincides with the short-range limit $y^2/2$
of Eq.~(\ref{COMLcor04c}) after introduction of the renormalized
effective Bohr radius
$\tilde{\lambda}=\lambda(1+\gamma)/\sqrt{1+2\gamma}$. Since
parameter $\gamma$ in this expression is usually less than or of
the order of unity, this renormalization is not significant, and
we can conclude that an approximation of the exciton wave function
by Eq.~(\ref{excwf}) gives reasonable results at least for the
short-range interface disorder. 
Collecting together all  results for the variance we obtain in the
limit of short-range interface disorder
\begin{equation}\label{U02_wn2D}
W^2=2h^2\left[1-f_{12}(L/\sigma_{\|})\right]\frac{\sigma_{\perp}^2}{\lambda^2}
\left[\frac{V_h^2\chi_h^4}{\beta_e^2}+8V_eV_h\chi_e^2\chi_h^2
 +\frac{V_e^2\chi_e^4}{\beta_h^2}\right].
\end{equation}
Apart from  the vertical-correlation factor  $1-f_{12}$ this
expression coincides with results obtained in
Ref.~\onlinecite{Zimmermann94a}. As expected, in the case when
holes have a significantly larger mass than electrons (the typical
situation for In$_{x}$Ga$_{1-x}$As/GaAs or
Al$_{x}$Ga$_{1-x}$As/GaAs quantum wells), the hole-hole term in
square brackets of Eq.~(\ref{U02_wn2D}) dominates. In the opposite
limit of the long-range interface correlations the result is
\begin{equation}\label{U02_gauss_as}
W^2=2h^2\left[1-f_{12}(L/\sigma_{\|})\right] \left(V_h\chi_h^2 +
V_e\chi_e^2\right)^2,
\end{equation}
which agrees with the conclusion of Ref.~\onlinecite{Castella98}
obtained for a different model of the interface disorder: in the
regime of the long-range correlations the distribution of the
effective potential reproduces the distribution of the interface
roughness. 
\section{Comparison with experimental results}
In order to compare calculations of $W$ with experimental
absorption spectra one needs to evaluate dynamics of excitons in a
random potential with given correlation properties. This problem
was intensively discussed in
literature\cite{Efros83a,EfrWetWor95,EfrosRaikh88,Runge02a} and we
are going to use the results of the cited papers in conjunction
with our analysis of the effective potential. There are two main
models of exciton dynamics in a random model. One of them treats
excitons quantum-mechanically in the limit of negative and large
energies, while describing a most important intermediate region
using an interpolation procedure\cite{Efros83a, EfrWetWor95}. In
this approach the absorption line has an asymmetric shape with the
linewidth proportional to $W^2$. In the second approach excitons
are treated semiclassically\cite{EfrosRaikh88}; if the underlying
compositional or interface roughness disorder is described by the
Gaussian random process, then the shape of the exciton line is
also approximately Gaussian with FWHM equal to
$\Delta=2\sqrt{2\ln(2)}W$. A transition between quasi-classical
and quantum regimes of exciton dynamics is determined by a
parameter $\nu=W/K_c$. Here $K_c=\hbar^2/2M\ell_c^2$ is a kinetic
energy of an exciton confined in a spatial region of size
$\ell_c$, where $\ell_c$ is a suitably defined correlation length
of the random potential\cite{Runge02a} (also see Appendix). The
quantum limit corresponds to the case $\nu\lesssim 1$, while the
semiclassical approximation is valid when $\nu\gg 1$.

\begin{figure}[bp]
\includegraphics{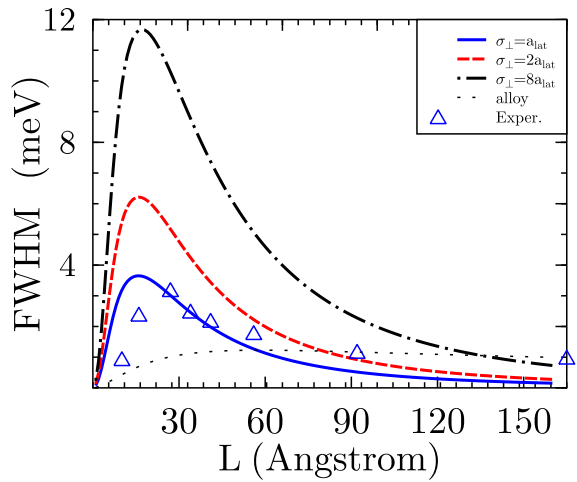}
 \caption{Dependence of the interface roughness induced broadening on the perpendicular correlation length
 (``island size") $\sigma_{\perp}$. The correlation length is given in terms of a number of lattice
 constants  ($a_{lat}=5.869\AA$). For all curves the
 composition concentration is $x=0.12$, and the vertical correlation length parameter
 is fixed by $\sigma_{\parallel}=a_{lat}$.  The dotted line is the rescaled alloy disorder contribution. The
 experimental data, shown by triangles, are taken from
 Ref.~\onlinecite{Bertolet88} and are presented here for comparison
 only.} \label{fig_perp}
\end{figure}

\begin{figure}[tbp]
\includegraphics{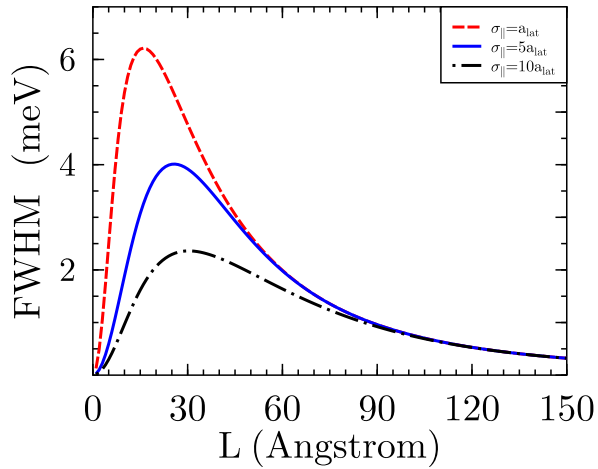}
 \caption{Dependence of the interface broadening on the vertical correlation length
 (``island size") $\sigma_{\parallel}$. All curves are drawn for $\sigma_{\perp}=2a_{lat}$.
 } \label{fig_vert}
\end{figure}

In order to compare results obtained here with optical
experimental data, we will make use of the semiclassical theory of
exciton absorption, which according to Ref.~\onlinecite{Runge02a}
can be reliably applied to the situation under consideration
(typical material parameters for an In$_{x}$Ga$_{1-x}$As/GaAs QW
yield $\nu\sim 5$ for $\sigma_{\perp}=2a_{lat}$, and $\nu$
increases with increase of the correlation length).
 Using the semiclassical relation for $\Delta$ we plot the dependence
of the interface roughness contribution to the exciton line width
as a function of the well width in Figs.~\ref{fig_perp} and
\ref{fig_vert}. Figure~\ref{fig_perp} represents curves for
different lateral correlation lengths (``island sizes"), while
Fig.~\ref{fig_vert} shows how these results are modified by the
inter-wall correlations. The first thing to notice is that the
changes in the lateral correlation length affect the height but
not the position of the maximum and the shape of the curve.
Comparison with the experimental data reproduced in this figure
shows that an increase in $\sigma_{\perp}$ drives the curves away
from the experimental results. This has to be compared with the
results of incorporating the inter-wall correlations,
Fig.~\ref{fig_vert}. An increase in $\sigma_{\|}$ not only
significantly reduces the height of the curve maximum, but also
shifts its position toward larger values of $L$, and widens it.

\begin{figure}[tbp]
\includegraphics{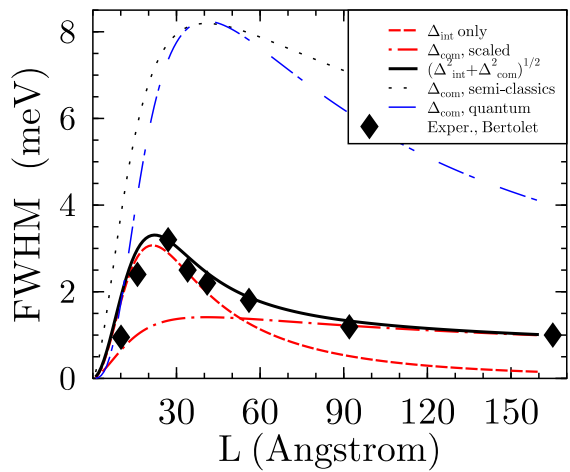}
 \caption{The thick solid line is the best fit for the total line-width displayed together with experimental results
 (triangles) from  Ref.~\onlinecite{Bertolet88}. The separate
 contributions of the interface disorder ($\Delta_{int}$) and of
 the rescaled alloy disorder ($\Delta_{alloy}$) are also shown.
 The parameter-free theory for alloy disorder contribution discussed in Ref.~\onlinecite{EfrWetWor95}
 is shown by the dashed-dotted line (without adiabatic approximation). The semiclassical limit of
 the alloy disorder contribution\cite{Zimmermann97a} is shown by the dotted line. The fitted parameters
 are $\sigma_{\parallel}=3\,a_{lat},\ \sigma_{\perp}=1\,a_{lat}$.
 } \label{fig_fit}
\end{figure}

It would be interesting to try to fit the experimental data
presented in Fig.~\ref{fig_exp} with the results of our
calculations. To this purpose one also needs to know the
contribution of alloy disorder to the total line width. Making use
of the model of short-range compositional disorder in a
QW\cite{EfrWetWor95} without the adiabatic approximation, one can
obtain the following formula for the alloy disorder induced
variance\cite{Zimmermann97a}:
\begin{equation}\label{alloydis}
  \frac{a_{lat}^3x(1-x)}{8\pi\lambda^2}\left[
  \frac{\alpha_h^2}{\beta_e^2}\int_{-L/2}^{L/2}\chi_h(z)^4\,dz+
  8\alpha_h\alpha_e\int_{-L/2}^{L/2}\chi_h(z)^2\chi_e(z)^2\,dz+
  \frac{\alpha_e^2}{\beta_h^2}\int_{-L/2}^{L/2}\chi_e(z)^4\,dz\right],
\end{equation}
where $a_{lat}$ is the lattice constant, and
$\alpha_{e,h}=dV_{e,h}/dx$ characterize the rate of the shift of
conduction and valence bands with composition, $x$. The formula is
given for the case of a QW made of a ternary alloy. The
semiclassical theory of the exciton line width again yields :
$\Delta=2\sqrt{2\ln(2)}W$, while the interpolation procedure for
the quantum limit gives $\Delta\approx 0.59MW^2$. The
In$_{0.12}$Ga$_{0.88}$As/GaAs QW is  intermediate between these
two limits, since $\nu\sim 1$. In Fig.~\ref{fig_fit} we present
both $\Delta$ dependencies on well thickness. Unfortunately, as we
can see, there exists a strong discrepancy between theoretical
estimates of the contribution from alloy disorder and experimental
results. It is not the goal of this paper to uncover the causes of
this discrepancy. However, common wisdom tells us that in the bulk
limit of very wide QW ($L>a_B$) only  alloy disorder contribution
should survive. The simplest way to adjust the theory is to
introduce a phenomenological scaling down of $\Delta_{comp}$ to
the value that should coincide with experimental results in the
limit of large $L$ asymptote. Although at present the reason for
such re-scaling is unknown, it is hard to imagine that the proper
theory of alloy disorder contribution will change the dependence
of the variance on well thickness, $L$, determined by the
integrals in Eq.~(\ref{alloydis}). Since our main purpose is to
elucidate the role of the inter-wall correlations rather than to
revise existing theories of the alloy disorder, we carry out this
operation keeping in mind its purely technical nature. The results
of the best fit performed in this way are shown in
Fig.~\ref{fig_fit} along with the best fit values of lateral and
vertical correlation lengths. Performing a similar fitting
procedure for other experimental dependencies of FWHM on well
thickness shown in Fig.~\ref{fig_exp} we obtained the following
values (normalized on lattice constant $a_{lat}$) for lateral and
inter-wall correlation lengths in In$_{x}$Ga$_{1-x}$As/GaAs QWs:
for $x=0.09$ $\sigma_{\bot}=1,\ \sigma_{\|}=20$, for $x=0.18$
$\sigma_{\bot}=1,\ \sigma_{\|}=10$, for $x=0.11$
$\sigma_{\bot}=1,\ \sigma_{\|}=5$, for $x=0.12$ $\sigma_{\bot}=1,\
\sigma_{\|}=3$, for $x=0.135$ $\sigma_{\bot}=4,\ \sigma_{\|}=13$.
 The most important result of
this exercise is the demonstration that the fit would not be
possible at all without taking into account the inter-wall
correlations. We would also like to stress that it is not possible
to achieve a good agreement with experimental results by omitting
the inter-wall correlations, and using only alloy disorder scaling
as an additional fitting parameter. We conclude, therefore, that
the inter-wall correlations play an important role and must be
taken into account when interpreting experimental results.


\section{Conclusion}
In conclusion, in this work we address the influence of vertical
inter-wall correlations between rough interfaces on the exciton
line shape. We show that the presence of these correlations
strongly suppresses the interface disorder contribution into
inhomogeneous broadening. The latter means that for narrow quantum
wells it might happen that the exciton line width tells more about
the quality of barrier material  than about the quality of
interface contrary to what is often claimed in the experimental
literature. On the other hand, the differences in inter-wall
correlation lengths can account for the variety of positions,
strengths and sharpness of FWHM dependence on the well width for
experimental data, obtained by different research groups.

\begin{acknowledgments}
We are grateful to S. Schwarz for reading and commenting on the
manuscript. The work is supported by AFOSR grant F49620-02-1-0305
and PSC-CUNY Research grants.
\end{acknowledgments}

\appendix
\section{\label{sec:Ap1} Dependence of the lateral correlation function on $R$}
\begin{figure}[tbp]
\includegraphics{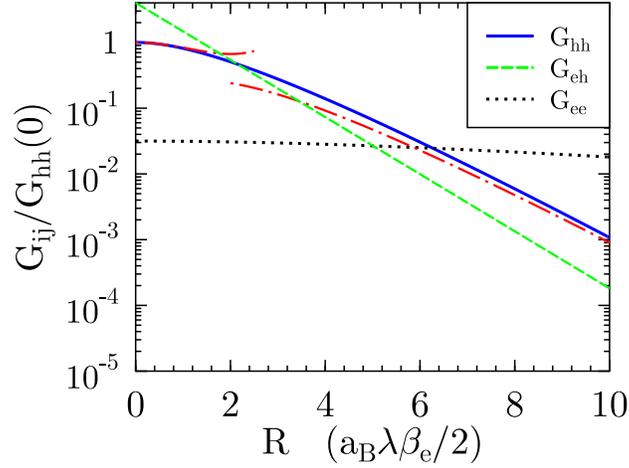}
 \caption{Semi-log plot of the normalized lateral correlation functions
 $G_{ij}(R;y_j,\alpha)$. The normalization is chosen in a way
 to have the hole-hole correlator equal to unity at $R=0$. Data
 are given for $y_h=0.1$. The ratio $m_e/m_h=0.178$ was chosen to depict the realistic case
 of InGaAs/GaAs QW.
 The dotted-dashed lines for hole-hole correlator are limits of
 small and large $\tilde{R}$ [See Eq.~(\ref{GRee})] }\label{fig3a}
\end{figure}

Calculation of the exciton absorption line shape is equivalent in
the dipole approximation to the estimation of the optical density
(OD) function\cite{EfrosRaikh88}:
\begin{equation}\label{ODfun}
  A(\varepsilon)=\left\langle\sum_i\left|\int d^2\,R
  \psi_i(\textbf{R})\right|^2\delta(\varepsilon-\varepsilon_i)\right\rangle,
\end{equation}
where $\varepsilon_i$ and $\psi_i(R)$ are corresponding energy and
wave function of the Schr\"{o}dinger equation (\ref{SEforCOM}) for
exciton COM. The shape of $A(\varepsilon)$ depends on the strength
of disorder. The latter can be roughly measured by parameter
$\nu=W/K_c$, where $W=\sqrt{\left\langle U_{eff}(R)^2
\right\rangle}$ is a variance of the potential energy induced by
fluctuations and $K_c=\hbar^2/2M\ell_c^2$ is the kinetic
("correlational") energy of the exciton. The parameter $\ell_c$
determines the confinement of the exciton COM wave function. It
can be extracted from  knowledge\cite{Runge02a} of the lateral
dependence on $R$ for the correlation functions
$G_{ij}(R,y_j,\alpha)$ since
\begin{equation}\label{lcc}
\ell_c^D=\int\, d^DR' \langle
U({\bf{R}})U({\bf{R-R'}})\rangle/W^2.
\end{equation}
 For the ``white-noise" height-height correlator (\ref{whitenoise})  one
can show that all distances are scaled by factors
$(\lambda\beta_i)/2$, i.e. $\tilde{R}=2R/(\lambda\beta_i)$. For
the cross-term correlator in the limit of heavy holes and light
electrons the result is simple again
\begin{equation}\label{GReh}
  G_{eh}(R;y_h,\alpha)\approx G_{eh}(R;y_h,0) =2y_h^2\exp(-\tilde{R}),
\end{equation}
while for the diagonal terms we have $G_{ii}(R;y_j,0)=y_j^2/2
f(\tilde{R}_i)$, where
\begin{equation}\label{GReh2}
  f(\tilde{R})=\frac{4}{\pi}
  \int_0^{\pi}d\theta\int_0^{\infty}d\rho\,\rho
  \exp\left(-\rho-\sqrt{\tilde{R}^2-2\rho\tilde{R}\cos\theta+\rho^2} \right).
\end{equation}
Function $f(\tilde{R})$ has the following limits for small and
large distances:
\begin{equation}\label{GRee}
f(\tilde{R})\approx \left\{
\begin{array}
[c]{cc}%
1-\frac{1}{4}\tilde{R}^2+\frac{1}{12}\tilde{R}^3, & \tilde{R}< 1\\
\sqrt{\frac{\pi\tilde{R}^3}{8}}\exp(-\tilde{R}), & \tilde{R} \gg
1.
\end{array}
\right.
\end{equation}
 Thus, even though the initial correlator of interface
 fluctuations was of the white noise type, the effective noise for the exciton potential is colored
 with exponential tails and with a correlation length of the order of the exciton radius.
 Similar results were obtained earlier for the bulk compositional
 fluctuations\cite{Lozovik85} and for the island model of interfacial
 roughness\cite{Castella98}. One can show that such exponential asymptotes also persist for the
long-range Gaussian correlator, Eq.~(\ref{gausscor}), when $R/y\gg
1$. Since
\begin{equation}\label{GReevsGRhh}
G_{ee}(R)=\left(\frac{m_e}{m_h}\right)^2G_{hh}\left(R
\frac{m_e}{m_h}\right),
\end{equation}
we can see that the lateral part of the electron-electron
correlation function is suppressed by the factor $(m_e/m_h)^2$ but
has larger correlation length. Overall, the localization length,
$\ell_c$, is determined however by the hole-hole term: $\ell_c\sim
\lambda\beta_e$.
 The normalized
lateral correlation functions $G_{ij}(R)$ are shown in Fig.~%
\ref{fig3a}.


\end{document}